\RequirePackage{ifpdf}
\ifpdf % We are running pdfTeX in pdf mode
\documentclass[pdftex]{sigma}
\else
\documentclass{sigma}
\fi

\def\DIS{\displaystyle}

\def\Z{{\mathbb Z}}

\def\F{{\mathbb F}}

\def\bm#1{\mbox{\boldmath $#1$}}
\def\fbm#1{\mbox{\scriptsize \boldmath $#1$}}

\begin{document}

\allowdisplaybreaks

\renewcommand{\thefootnote}{$\star$}

\renewcommand{\PaperNumber}{027}

\FirstPageHeading

\ShortArticleName{Two Point Correlation Functions for a Periodic Box-Ball System}

\ArticleName{Two Point Correlation Functions\\ for a Periodic Box-Ball System\footnote{This
paper is a contribution to the Proceedings of the Conference ``Integrable Systems and Geomet\-ry'' (August 12--17, 2010, Pondicherry University, Puducherry, India). The full collection is available at \href{http://www.emis.de/journals/SIGMA/ISG2010.html}{http://www.emis.de/journals/SIGMA/ISG2010.html}}}

\Author{Jun MADA~$^\dag$ and Tetsuji TOKIHIRO~$^\ddag$}

\AuthorNameForHeading{J.~Mada and T.~Tokihiro}

\Address{$^\dag$~College of Industrial Technology,     Nihon University,\\
\hphantom{$^\dag$}~2-11-1 Shin-ei, Narashino, Chiba 275-8576, Japan}
\EmailD{\href{mailto:mada.jun@nihon-u.ac.jp}{mada.jun@nihon-u.ac.jp}}

\Address{$^\ddag$~Graduate School of Mathematical Sciences,    University of Tokyo,\\
\hphantom{$^\ddag$}~3-8-1 Komaba, Tokyo 153-8914, Japan}
\EmailD{\href{mailto:toki@ms.u-tokyo.ac.jp}{toki@ms.u-tokyo.ac.jp}}

\ArticleDates{Received December 13, 2010, in f\/inal form March 02, 2011;  Published online March 21, 2011}

\Abstract{We investigate correlation functions in a periodic box-ball system.
For the second and the third nearest neighbor correlation functions,
we give explicit formulae obtained by combinatorial methods.
A recursion formula for a specif\/ic $N$-point functions is also presented.}


\Classification{37B15; 37K10; 81R12; 82B20}

\section{Introduction}
\label{sec:introduction}

A periodic box-ball system (PBBS) is a soliton cellular automaton
obtained by ultradiscretizing the KdV equation~\cite{TS, YT}.
It can also be obtained at the $q \to 0$ limit
of an integrable lattice model (a generalised 6 vertex model)~\cite{FOY, HHIKTT}.
Let $V_1$ be a 2-dimensional complex vector space with basis $\vec{e}_0$ and $\vec{e}_1$.
If we consider $N$ tensor product space of $V_1$, $V:=V_1^{\otimes N}$,
the transfer matrix of the generalised 6 vertex model is a map (endmorphism) $T_k(x;q):  V \to V$
with a~spectral parameter $x$ and a deformation parameter $q$.
The positive integer $k$ denotes that the dimension of the auxiliary vector space is $k+1$.
An important property of the transfer matrices is their commutativity:
$T_k(u;q)T_l(v;q)=T_l(v;q)T_k(u;q)$ for arbitrary $k$, $l$, $u$, $v$.
Hence they consist a complete set of diagonal operators and the lattice models are integrable.
It is also noted that the Hamiltonian of the quantum $XXZ$ spin model is essentially given by
$
H_{XXZ}:=\lim\limits_{x \to 1}\frac{\partial}{\partial x} \log T_1(x;q).
$
If we take a limit $ \lim\limits_{q \to 0,\, x \to 1}T(x;q)=: T$, the transfer matrix $T$
maps each monomial
$
 |i_1i_2\cdots i_N\rangle :=\vec{e}_{i_1}\otimes \vec{e}_{i_2} \otimes \cdots \otimes \vec{e}_{i_n} \in V
$
to a monomial.
By identifying $|i_1i_2\cdots i_N\rangle$ with a 10 sequence $i_1i_2\cdots i_N$, $T$ gives the time evolution
of the PBBS.
Using this relation, we can obtain several important properties of the PBBS such as the conserved quantities, a relation to the string hypothesis, a fundamental cycle and so on.

One of the main problems of quantum integrable systems now is to obtain
correlation functions which is fairly dif\/f\/icult even for the $XXZ$ model
and the 6 vertex model~\cite{JM}.
Since a~correlation function of the $XXZ$ model or the 6 vertex model is,
roughly speaking, transformed to a correlation function of the PBBS in the limit $q \to 0$.
For example, a two point function $\langle s_i^z s_j^z \rangle$ of the $XXZ$ spin model is transformed to
the probability that both $i$th and $j$th componets of the PBBS are $1$.
Hence, from the view point of integrable lattice models or quantum integrable models,
it may actually give some new insights into the correlation functions
of the vertex models themselves to obtain correlation functions of the PBBS.

In~\cite{MT2010},
we gave expressions for $N$-point functions using the solution for the PBBS
expressed in terms of the ultradiscrete theta functions.
We also gave expressions for the one point, the nearest and the second nearest neighbor correlation functions.
Note that the correlation functions def\/ined there and in this article are not the direct $q \to 0$ limit of
the correlation functions of the corresponding lattice model; such a limit generally gives trivial results \cite{MT2010}.
We def\/ine the correlation functions on the phase space of a set of f\/ixed conserved quantities.
In this article, we show that the second nearest neighbor correlation function can be simplif\/ied with some combinatorial formulae,
and give an explicit formula for the third nearest neighbor correlation function by combinatorial methods.
A recursion formula for a specif\/ic $N$-point function is also presented.

\begin{figure}[t]
\centering
 \includegraphics[width=.5\linewidth]{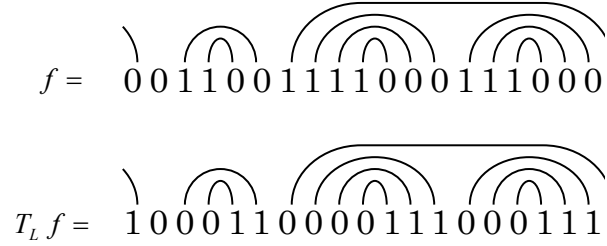}
 \caption{Def\/inition of $T_L$ for $f\in\Omega_L$.}
 \label{fig:TL}
\end{figure}

The PBBS can be def\/ined in the following way.
Let $L\ge3$ and let $\DIS \Omega_L=\big\{\,f\,|\,f:[L]\to\{0,1\}\ \mbox{such that}\
\sharp f^{-1}(\{1\})<L/2\,\big\}$ where $[L]=\{1,2,\ldots,L\}$.
When $f\in \Omega_L$ is represented as a~sequence of $0$s and $1$s, we write
\[
 f(1)f(2)\cdots f(L).
\]
The mapping $T_L:\Omega_L\to \Omega_L$ is def\/ined as follows (see Fig.~\ref{fig:TL}):
\begin{enumerate}\itemsep=0pt
\item In the sequence $f$ f\/ind a pair of positions $n$ and $n+1$
      such that $f(n)=1$ and $f(n+1)=0$, and mark them;
      repeat the same procedure until all such pairs are marked.
      Note that we always use the convention that the position $n$ is def\/ined in $[L]$,
       i.e.~$n+L\equiv n$.
\item Skipping the marked positions we get a subsequence of $f$;
      for this subsequence repeat the same process of marking positions,
      so that we get another marked subsequence.
\item Repeat part 2 until one obtains a subsequence consisting only of $0$s.
      A typical situation is depicted in Fig.~\ref{fig:TL}.
      After these preparatory processes,
      change all values at the marked positions simultaneously;
      One thus obtains the sequence $T_Lf$.
\end{enumerate}
The pair $(\Omega_L,T_L)$ is called a PBBS of length $L$~\cite{YT,MIT2006}.
An element of $\Omega_L$ is called a state, and the mapping $T_L$ the time evolution.

The conserved quantities of the PBBS are def\/ined as follows.
Let $Q_j(f)$ be the number of $10$ pairs in $f$ marked at $j$th step in the def\/inition of the mapping $T_L$.
Then we obtain a~nonincreasing sequence of
positive integers, $Q_j(f)$ $(j=1,2,\ldots,m)$.
This sequence is conserved in time, that is,
\[
Q_j(f)=Q_j(T_Lf)\equiv Q_j\qquad (j=1,2,\ldots,m).
\]
For example, $(Q_1,Q_2,Q_3,Q_4)\!=\!(3,3,2,1)$ for $f$ given in Fig.~\ref{fig:TL}.
As the sequence $(Q_1,Q_2,\ldots,Q_m)$ is nonincreasing,
we can associate a Young diagram with it by regarding $Q_j$ as the number of squares in the $j$th column of the diagram.
The lengths of the rows are also weakly decreasing positive integers.
Let the distinct row lengths be $P_1>P_2>\cdots>P_\ell$ and let $n_j$ be the number of times that the length $P_j$
appears. The set $\{P_j,n_j\}_{j=1}^\ell$ is another expression for the conserved quantities of the PBBS.

First we summarize some useful properties of the PBBS.
We say that $f$ has (or that there is) a $10$-wall at position $n$
if $f(n)=1$ and $f(n+1)=0$. Let the number of the $10$-walls be $s$
and the positions be denoted by $a_1<a_2<\cdots<a_s$.

We introduce a procedure called $10$-insertion.
For $(b_1,b_2,\ldots,b_L)\in \Z_{\ge0}\times \Z_{\ge0}\times \cdots \times\Z_{\ge0}\ (\equiv(\Z_{\ge0})^L)$ and $d:=\sum\limits_{i=1}^L b_i$,
the $10$-insertion is def\/ined as $I(b_1,b_2,\ldots,b_L):=I_2(b_1,b_2,\ldots,b_L)\circ I_1$, where $I_1 :
\Omega_L\to \Omega_{L+2s}$ is the mapping:
\[
 (I_1f)(n) =\left\{ \begin{array}{ll}
  1 & (n=a_j+2j-1,\ j\in[s]), \\[1mm]
  0 & (n=a_j+2j,\ j\in[s]), \\[1mm]
  f(n) & (1\le n\le a_1), \\[1mm]
  f(n-2j) & (a_j+2j+1\le n\le a_{j+1}+2j,\ j\in[s-1]), \\[1mm]
  f(n-2s) & (a_s+2s+1\le n\le L+2s) \\
 \end{array} \right.
\]
and the mapping $I_2(b_1,b_2,\ldots,b_L):\Omega_{L+2s}\to\Omega_{L+2(d+s)}$ is def\/ined to be
\begin{gather*}
  (I_2(b_1,b_2,\ldots,b_L)f')(n)
  =\left\{ \begin{array}{ll}
  f'(n-2\hat{b}_{j-1}) & (n=j+2\hat{b}_{j-1},\ j\in[L]), \\ [1mm]
  1 & (n=j+2\hat{b}_{j-1} +2h-1,\ j\in[L],\ h\in[b_j]), \\[1mm]
  0 & (n=j+2\hat{b}_{j-1} +2h,\ j\in[L],\ h\in[b_j]), \\
 \end{array} \right.
\end{gather*}
where $f'\equiv I_1f\in\Omega_{L+2s}$, $b_0=0$ and $\hat{b}_j :=\sharp \big\{i\in[s]\,\big|\,a_i\le j\big\}+\sum_{i=1}^j b_i$.
For example,
\begin{alignat*}{4}
& f && =\, && 0011100111000001101000111000000,& \nonumber \\
& I_1f && =\, && 00111\fbox{\hskip-1.2mm 10\hskip-1.1mm}0011
 1\fbox{\hskip-1.2mm 10\hskip-1.1mm}0000011
 \fbox{\hskip-1.2mm 10\hskip-1.1mm}01\fbox{\hskip-1.2mm 10\hskip-1.1mm}000
 111\fbox{\hskip-1.2mm 10\hskip-1.1mm}000000, &  \\
&  I(1_9,1_{22},1_{29})f && =\,  && 00111\fbox{\hskip-1.2mm 10\hskip-1.1mm}0011
 \underline{10}1\fbox{\hskip-1.2mm 10\hskip-1.1mm}0000011
 \fbox{\hskip-1.2mm 10\hskip-1.1mm}01\fbox{\hskip-1.2mm 10\hskip-1.1mm}000
 \underline{10}111\fbox{\hskip-1.2mm 10\hskip-1.1mm}0000\underline{10}00, &
\end{alignat*}
where $(1_9,1_{22},1_{29})$ denotes the $10$ sequence whose $9$, $22$, and $29$th elements are $1$ and the others are $0$, that is,
\[
(1_9,1_{22},1_{29}) \equiv (0,0,0,0,0,0,0,0,1,0,0,0,0,0,0,0,0,0,0,0,0,1,0,0,0,0,1,0,0,0,0),
\]
and the expressions $\fbox{\hskip-1.2mm 10\hskip-1.1mm}$ and $\underline{10}$ denote the $10$s inserted at $f\mapsto I_1f$
and at $I_1f\mapsto I_2(b_1,b_2,\ldots$, $b_L)(I_1f)$ respectively.

\section{Correlation functions of PBBS}
\label{sec:cy12n}

We consider the PBBS with $L$ boxes and $M$ balls.
As shown in~\cite{MT2010}, the correlation functions are def\/ined on the states with the same conserved quantities.
Let $Y$ be the Young diagram corresponding to the partition of $M$:
\[
\big(\underbrace{P_1,P_1,\dots,P_1}_{n_1},\underbrace{P_2,P_2,\dots,P_2}_{n_2},
\ldots,\underbrace{P_\ell,P_\ell,\dots,P_\ell}_{n_\ell}\big),
\]
where $P_1>P_2>\cdots >P_\ell\ge1$ and $\sum\limits_{j=1}^\ell n_jP_j=M$.
The number of squares in the $j$th largest raw of $Y$ is $P_j$.
In the PBBS, the conserved quantities are characterized by the Young diagram~$Y$~\cite{ToTaS}.
When we denote by $\Omega_Y$ the set of states with conserved quantities given by~$Y$,
the $N$-point function of the PBBS is def\/ined as
\[
\langle s_1,s_2,\ldots,s_N \rangle_Y
=\frac{1}{|\Omega_Y|} \sum_{f\in \Omega_Y} f(s_1)f(s_2)\cdots f(s_N).
\]
Since the $N$-point function $\langle s_1,s_{1}+d_1,\ldots,s_{1}+d_{N-1} \rangle_Y$ does not
depend on the specif\/ic site $s_1$
(because of translational symmetry), we denote
\[
 C_Y(d_1,d_2,\ldots,d_{N-1})\equiv
 \langle s_1,s_{1}+d_1,\ldots,s_{1}+d_{N-1} \rangle_Y,
\]
where $1\le d_1<d_2<\cdots<d_{N-1}< L$.
The $1$-point function $C_Y(\varnothing):=\langle s_1 \rangle_Y$ and the $2$-point function $C_Y(1):=\langle s_1, s_1+1\rangle_Y$
are easily calculated as $  C_Y(\varnothing)=\frac{M}{L} $ and  $  C_Y(1) =\frac{M-s}{L}$ where $s=\sum\limits_{i=1}^\ell n_i$.

Although general $N$-point functions are dif\/f\/icult to evaluate by elementary combinatorial methods, the correlation functions
$C_Y(1,2,\ldots,N-1)=\langle s_1,s_1+1,s_1+2,\dots,s_1+N-1 \rangle_Y$ enjoy a simple recursion formula.
Let us put
\begin{gather*}
 k_i  := \left\{ \begin{array}{ll}
 n_j & (i=P_j), \\[1mm]
 0 & \mbox{otherwise,} \\
 \end{array} \right. \qquad
 \hat{k}_i  := \sum_{j=i}^{P_1} k_j, \qquad
 L_0  :=L,\qquad L_i:= L_{i-1}-2\hat{k}_i,
\end{gather*}
for $i=1,2,\ldots,P_1$.
Note that $k_i$ and $\hat{k}_i$ are the number of the rows in $Y$ with the length $i$ and that with the length greater than or equal to $i$ respectively.
For the Young diagram $Y$, let $Y_i$ be the Young diagram corresponding to the partition
\begin{gather*}
  \Big(\underbrace{\max \big[ P_1-i,0\big],\max \big[ P_1-i,0\big],\dots,\max \big[ P_1-i,0\big]}_{n_1}, \\
  \qquad \underbrace{\max \big[ P_2-i,0\big],\max \big[ P_2-i,0\big],\dots,\max \big[ P_2-i,0\big]}_{n_2},\ldots \\
  \qquad \quad \ldots,\underbrace{\max \big[ P_\ell-i,0\big],\max \big[ P_\ell-i,0\big],\dots,\max \big[ P_\ell-i,0\big]}_{n_\ell}\Big).
\end{gather*}
The following two lemmas immediately follow from the def\/inition of $I_1$ and the procedure of $10$-insertion.

\begin{lemma}
\label{lemma1}
In a state $f=I_1f_1$ $(f_1\in \Omega_{Y_1})$, no pattern with $101$ exists.
\end{lemma}

\begin{lemma}
\label{lemma2}
The number of the pattern $11$ in $f_1\in \Omega_{Y_1}$ is equal to that of the pattern $111$ in $I_1 f_1$, and it does not depend on the specific state $f_1$.
If we denote it by ${\cal N}_2$, it is calculated as
\begin{equation}
{\cal N}_2=\sum_{i=3}^{P_1} k_i(i-2).
\label{calN2}
\end{equation}
\end{lemma}
Then we have the following theorem.

\begin{theorem}
\label{Prop3}
\begin{equation*}
C_Y(1,2,\ldots,r+1)=\left( \frac{L_1}{L}\prod_{i=1}^r\frac{(L_1-i)}{(L_1+k_1-i)}\right) C_{Y_1}(1,2,\ldots,r)\qquad (r=1,2,\dots).
%\label{Npointrecursion}
\end{equation*}
In particular,
\begin{gather*}
C_Y(1,2) =\frac{(L_1-1)}{L(L_1+k_1-1)}\sum_{i=3}^{P_1}k_i(i-2),\\
C_Y(1,2,3) =\frac{L_1(L_1-1)(L_1-2)}{L(L_1+k_1-1)(L_1+k_1-2)}C_{Y_1}(1,2).
\end{gather*}
\end{theorem}

\begin{proof}
We call a `block' for a repeated $10$ pattern of the form $\underline{10}\,\underline{10}\cdots\underline{10}$.
We also call a $111$ pattern for the pattern of the form $111$.

First let us consider the case $r=1$. From Lemma~\ref{lemma2}, the number of $111$ patterns in $I_1f_1$ is equal to $\sum\limits_{i=3}^{P_1} k_i(i-2)$.
Let ${\cal B}_1$ be a set $\big\{\,B_1\in (\Z_{\ge0})^{L_1}\,\big|\,\sum\limits_{i=1}^{L_1}b_i =k_1,\,B_1=(b_1,b_2,\ldots,b_{L_1})\,\big\}$.
In the process $I_1f_1 \mapsto I(B_1)f_1$, if $\underline{10}$ blocks is inserted between the left $1$ and the middle $1$ of a $111$ pattern,
then the $111$ pattern in $I(B_1)f_1$ disappears.
Let $\Psi_1:\Z_{\ge0}\times (\F_2\times \F_2)\to \{0,1\}$ be the function:
\[
\Psi_1(b;u_1,u_2) :=\left\{ \begin{array}{ll}
 1 & \big(b=0,\ (u_1,u_2)=(1,1)\big), \\[1mm]
 0 & \mbox{otherwise} \\
 \end{array} \right.
\]
and  $\psi_1(f_1) :=\sharp \big\{\,i\in[L_1]\,\big|\,\big(f_1(i),f_1(i+1)\big)=(1,1)\,\big\}$.
Since a state of $\Omega_Y$ is obtained by 10-insertion to a state in $\Omega_{Y_1}$,
we have
\begin{gather*}
 C_Y(1,2)  =\frac{1}{L|{\cal B}_1||\Omega_{Y_1}|}
 \sum_{B_1\in {\cal B}_1} \sum_{f_1\in \Omega_{Y_1}} \sum_{i=1}^L \big(I(B_1)f_1\big)(i)\ \big(I(B_1)f_1\big)(i+1)\ \big(I(B_1)f_1\big)(i+2) \\
\phantom{C_Y(1,2)}{} =\frac{1}{L|{\cal B}_1||\Omega_{Y_1}|}
 \sum_{B_1\in {\cal B}_1} \sum_{f_1\in \Omega_{Y_1}} \sum_{i=1}^{L_1} \Psi_1(b_i;f_1(i),f_1(i+1)) \\
\phantom{C_Y(1,2)}{} =\frac{1}{L|{\cal B}_1||\Omega_{Y_1}|}
 \sum_{B_1\in {\cal B}_1} \sum_{f_1\in \Omega_{Y_1}} \sum_{i=1}^{L_1} \Psi_1(b_1;f_1(i),f_1(i+1)) \\
\phantom{C_Y(1,2)}{} =\frac{1}{L|{\cal B}_1||\Omega_{Y_1}|}
 \sum_{B_1\in {\cal B}_1} \sum_{f_1\in \Omega_{Y_1}} \psi_1(f_1) \Psi_1(b_1;1,1) \\
\phantom{C_Y(1,2)}{} =\frac{1}{L|{\cal B}_1||\Omega_{Y_1}|}
 \sum_{f_1\in \Omega_{Y_1}} \psi_1(f_1) \sum_{B_1\in {\cal B}_1} \Psi_1(b_1;1,1) \\
\phantom{C_Y(1,2)}{} =\frac{{{L_1+k_1-2}\choose{k_1}}}{L|{\cal B}_1|} %\times
\left( \frac{1}{|\Omega_{Y_1}|}\sum_{f_1\in \Omega_{Y_1}} \psi_1(f_1) \right)  =\frac{(L_1-1)}{L(L_1+k_1-1)}\sum_{i=3}^{P_1}k_i(i-2).
\end{gather*}
From
\begin{equation}
\sum_{i=3}^{P_1} k_i(i-2) =\sum_{i=2}^{P_1} k_i(i-1) -\hat{k}_2 =L_1C_{Y_1}(1),
\label{cy1formula}
\end{equation}
we f\/ind
\[
C_Y(1,2) =\frac{L_1}{L}\frac{(L_1-1)}{(L_1+k_1-1)}C_{Y_1}(1).
\]

For general $r$, if we def\/ine
\begin{gather*}
 \Psi_r(b_1,b_2,\dots,b_r; u_1,u_2,\ldots,u_{r+1})\\
 \qquad :=
\left\{
\begin{array}{cl}
1&\quad\big( (b_1,b_2,\dots,b_r)=(0,0,\dots,0),\ (u_1,u_2,\ldots,u_{r+1})=(1,1,\ldots,1)\big),\\[1mm]
0&\quad\mbox{otherwise,}
\end{array}
\right.
\end{gather*}
and
$\psi_r(f_1) :=\sharp \big\{\,i\in[L_1]\,\big|\,\big(f_1(i),f_1(i+1),\ldots,f_1(i+r)\big)=(1,1,\ldots,1)\,\big\}$,
we can proceed in a~similar manner to the case $r=1$ as
\begin{gather*}
  C_Y(1,2,\ldots,r+1)
 =\frac{1}{L|{\cal B}_1||\Omega_{Y_1}|}
 \sum_{B_r\in {\cal B}_1} \sum_{f_1\in \Omega_{Y_1}} \sum_{i=1}^L \big(I(B_1)f_1\big)(i)\ \cdots \ \big(I(B_1)f_1\big)(i+r+1) \\
\phantom{C_Y(1,2,\ldots,r+1)}{}
 =\frac{1}{L|{\cal B}_1||\Omega_{Y_1}|}
 \sum_{B_r\in {\cal B}_1} \sum_{f_1\in \Omega_{Y_1}} \sum_{i=1}^{L_1} \Psi_r(b_i,\ldots,b_{i+r} ;\ f_1(i),\ldots,f_1(i+r+1)) \\
\phantom{C_Y(1,2,\ldots,r+1)}{}
 =\frac{1}{L|{\cal B}_1||\Omega_{Y_1}|}
 \sum_{B_r\in {\cal B}_1} \sum_{f_1\in \Omega_{Y_1}} \sum_{i=1}^{L_1} \Psi_r(b_1,\ldots,b_{r+1} ;\ f_1(i),\ldots,f_1(i+r+1)) \\
\phantom{C_Y(1,2,\ldots,r+1)}{} =\frac{1}{L|{\cal B}_1||\Omega_{Y_1}|}
 \sum_{B_1\in {\cal B}_1} \sum_{f_1\in \Omega_{Y_1}} \psi_r(f_1) \Psi_r(b_1,\cdots,b_{r+1};(1,1,\ldots,1)) \\
\phantom{C_Y(1,2,\ldots,r+1)}{}
 =\frac{{{L_1+k_1-r-1}\choose{k_1}}}{L|{\cal B}_1|} \times \left( \frac{1}{|\Omega_{Y_1}|}\sum_{f_1\in \Omega_{Y_1}} \psi_r(f_1) \right) \\
\phantom{C_Y(1,2,\ldots,r+1)}{}
 =\frac{{{L_1+k_1-r-1}\choose{k_1}}}{L{{L_1+k_1-1}\choose{k_1}}} \times L_1C_{Y_1}(1,2,\ldots,r)  \\
\phantom{C_Y(1,2,\ldots,r+1)}{}
 =\frac{L_1C_{Y_1}(1,2,\ldots,r)}{L}\prod_{i=1}^r\frac{(L_1-i)}{(L_1+k_1-i)},
\end{gather*}
which completes the proof.
\end{proof}

\section{The second and the third nearest neighbor\\ correlation functions}
\label{sec:combinatorial}

\subsection[Correlation function $C_Y(2)$]{Correlation function $\boldsymbol{C_Y(2)}$}
Using $10$-insertion and the Young diagram $Y_1$ def\/ined in the previous section,
we can rewrite $C_Y(2)$ as
\[
 C_Y(2) =\frac{1}{L|{\cal B}_1||\Omega_{Y_1}|}
 \sum_{B_1\in {\cal B}_1} \sum_{f_1\in \Omega_{Y_1}} \sum_{n=1}^L \big(I(B_1)f_1\big)(n)\ \big(I(B_1)f_1\big)(n+2).
\]
From this expression we can prove the following proposition which gives a key formula to calculate $C_Y(2)$.
\begin{proposition}
\label{Prop1}
Let $\Phi_0:\Z_{\ge0}\times (\F_2 \times \F_2) \to \Z_{\ge0}$ be the function defined by
\[
\Phi_0(b;u_1,u_2) :=\left\{ \begin{array}{ll}
 0 & (b=0), \\[1mm]
 b & \big(b\ne0,\ (u_1,u_2)=(0,1),(1,0)\big), \\[1mm]
 b-1 & \big(b\ne0,\ (u_1,u_2)=(0,0),(1,1)\big). \\
 \end{array} \right.
\]
Then we have
\begin{equation}
C_Y(2) =\frac{1}{L} ({\cal N}_2 +{\cal R}_2),
\label{CY2}
\end{equation}
where ${\cal N}_2$ is given in \eqref{calN2},
\[
{\cal R}_2 =\frac{1}{|{\cal B}_1|} \sum_{B_1\in {\cal B}_1} \sum_{\fbm{b}\in (\F_2)^2} \phi_0(f_1;{\bm b})\, \Phi_0(b_1;\bm{b})\qquad (f_1\in \Omega_{Y_1}),
\]
and
$\DIS \phi_0(f_1;{\bm b}) :=\sharp \big\{\,i\in[L_1]\,\big|\,\big(f_1(i),f_1(i+1)\big)=\bm{b}\,\big\}$.
\end{proposition}

\begin{proof}
We call a $1*1$ pattern for the pattern of the form $1*1$ where $*$ is either $1$ or $0$.
From Lemmas~\ref{lemma1} and \ref{lemma2}, we already know the number of $1*1$ patterns in $I_1 f_1$.
Hence we estimate the number of the $1*1$ patterns generated and eliminated by inserting blocks $\underline{10}\,\underline{10}\cdots\underline{10}$ in the process $I_1f_1\mapsto I(B_1)f_1$.
Since there appears no $1*1$ pattern between blocks, we have only to count the number of the patterns $101$ appeared in a block and a boundary and that of the patterns $111$ eliminated by $10$-insertion;
then, for $b_n\ge 1$, we observe the number of $1*1$ patterns appeared by the operation of $I_2(B_1)$ as
\begin{alignat*}{4}
& (1) \ \  &&  f_1(n)f_1(n+1)={\bf 00}\ \longrightarrow\ {\bf 0}\underline{10}\,\underline{10}\cdots\underline{10}{\bf 0}  \ && \Rightarrow\  +(b_n-1), & \\
& (2) \ \ && f_1(n)f_1(n+1)={\bf 10}\ \longrightarrow\ {\bf 1}\fbox{\hskip-1.2mm 10\hskip-1.1mm}\underline{10}\,\underline{10}\cdots\underline{10}{\bf 0} \ && \Rightarrow\ +b_n,&  \\
& (3) \ \ && f_1(n)f_1(n+1)={\bf 01}\ \longrightarrow\ {\bf 0}\underline{10}\,\underline{10}\cdots\underline{10}{\bf 1} \ && \Rightarrow\ +b_n, & \\
& (4) \ \ && f_1(n)f_1(n+1)={\bf 11}\ \longrightarrow\ {\bf 1}\underline{10}\,\underline{10}\cdots\underline{10}{\bf 1} \ &&  \Rightarrow\ +(b_n-1). &
\end{alignat*}
Here we write the referred $1,0$s in $I_1f_1$ in bold scripts (${\bf 1,0}$) for clarifying the results.
Note that if $f_1(n)f_1(n+1)={\bf 11}$ and $b_n=0$, then we have the patterns ${\bf 11}1$ or ${\bf 11}\fbox{\hskip-1.2mm 10\hskip-1.1mm}$ or ${\bf 11}\underline{10}$ after the operation of $I_2(B_1)$.
Hence, we have
\begin{gather*}
{\cal R}_2
 =\frac{1}{|{\cal B}_1||\Omega_{Y_1}|}
 \sum_{B_1\in {\cal B}_1} \sum_{f_1\in \Omega_{Y_1}}\!\!
\Big( \sharp \big\{ \mbox{`101'} \in I(B_1)f_1 \big\} -\sharp \big\{ \mbox{`111'} \in I_1f_1 \ \mbox{disappears in $I(B_1)f_1$} \big\} \Big) \\
\phantom{{\cal R}_2}{}
=\frac{1}{|{\cal B}_1||\Omega_{Y_1}|}
 \sum_{B_1\in {\cal B}_1} \sum_{f_1\in \Omega_{Y_1}} \sum_{i=1}^{L_1} \Phi_0(b_i;f_1(i),f_1(i+1)).
\end{gather*}
Here $\sharp \big\{ \mbox{`101'} \in I(B_1)f_1 \big\}$ denotes the number of `$101$'s in $I(B_1)f_1$ and $\sharp \big\{ \mbox{`111'} \in I_1f_1\ \mbox{disappears} $ $\mbox{in}\ I(B_1)f_1 \big\}$
denotes the number of `$101$'s which disappear by the operation $I_2(B_1)$.
Since
\[
(b_1,b_2,\ldots,b_{L_1}) \in {\cal{B}}_1 \ \longrightarrow \ (b_i,b_{i+1},\ldots,b_{L_1},b_1,\ldots,b_{i-1}) \in {\cal{B}}_1,
\]
we have
\[
\sum_{B_1\in {\cal B}_1} \Phi_0(b_i;f_1(i),f_1(i+1)) =\sum_{B_1\in {\cal B}_1} \Phi_0(b_1;f_1(i),f_1(i+1)).
\]
Thus we have
\begin{gather*}
\sum_{B_1\in {\cal B}_1}\sum_{i=1}^{L_1} \Phi_0(b_i;f_1(i),f_1(i+1))
 =\sum_{B_1\in {\cal B}_1}\sum_{i=1}^{L_1} \Phi_0(b_1;f_1(i),f_1(i+1)) \\
\hphantom{\sum_{B_1\in {\cal B}_1}\sum_{i=1}^{L_1} \Phi_0(b_i;f_1(i),f_1(i+1))}{}
=\sum_{B_1\in {\cal B}_1}\sum_{\fbm{b}\in (\F_2)^2} \phi_0(f_1;{\bm b})\, \Phi_0(b_1,\bm{b}).
\end{gather*}
Furthermore, if $f$ and $f'$ are the states of PBBS with the same number of boxes, balls and solitons, then $\phi_0(f;{\bm b})=\phi_0(f';{\bm b})$ for any ${\bm b} \in (\F_2)^2$,
in particular, $\phi_0(f_1;{\bm b}) =\phi_0(f_1';{\bm b}) $ for $f_1, f_1' \in \Omega_{Y_1}$.
Therefore we obtain~\eqref{CY2}.
\end{proof}

For the evaluation of $C_Y(2)$, we use the identity for binomial coef\/f\/icients given in the following lemma.
\begin{lemma}
For $k \le l \le n$, it holds that
\begin{equation}
{n \choose l}=\sum_{m=0}^{n-l} \left\{{k+m \choose k}-{k+m-1 \choose k} \right\}{n-k-m \choose l-k}.
\label{binomialidentity}
\end{equation}
\end{lemma}

\begin{theorem}
\label{Theorem1}
\begin{gather}
 C_Y(2)  =\frac{1}{L}\left( \sum_{i=3}^{P_1} k_i(i-2) +\frac{k_1(k_1-1)}{L_1+k_1-2} +\frac{2\hat{k}_2k_1}{L_1+k_1-1} \right)
\label{cy2}\\
\phantom{C_Y(2)}{}
=\frac{1}{L}\left( L_1C_{Y_1}(1) +\frac{k_1(k_1-1)}{L_1+k_1-2} +\frac{2\hat{k}_2k_1}{L_1+k_1-1} \right).
\label{cy21}
\end{gather}
\end{theorem}

\begin{proof}
There are ${L_1+k_1-m-2}\choose{k_1-m}$ elements with $b_1=m$ $(m \ge 1)$ in the set ${\cal B}_1$.
Since  $\phi_0(f_1;(0,1))$ $=\phi_0(f_1;(1,0))=\hat{k}_2 $ and $\phi_0(f_1;(0,0))+\phi_0(f_1;(1,1))=L_1-2\hat{k}_2$,
we have
\begin{gather*}
   \sum_{B_1\in {\cal B}_1} \sum_{\fbm{b}\in (\F_2)^2} \phi_0(f_1;{\bm b})\, \Phi_0(b_1;\bm{b})
 =\sum_{m=1}^{k_1} {{L_1+k_1-m-2}\choose{k_1-m}} \sum_{\fbm{b}\in (\F_2)^2} \phi_0(f_1;{\bm b})\, \Phi_0(m;\bm{b}) \\
  \qquad
 =\sum_{m=1}^{k_1} {{L_1+k_1-m-2}\choose{k_1-m}} \left( \sum_{\fbm{b}=(0,0),\,(1,1)} (m-1)\, \phi_0(f_1;{\bm b})
  +\sum_{\fbm{b}=(0,1),\,(1,0)} m\,\phi_0(f_1;{\bm b}) \right) \\
  \qquad
 =L_1 \sum_{m=1}^{k_1} (m-1){{L_1+k_1-m-2}\choose{k_1-m}} +2\hat{k}_2 {{L_1+k_1-2}\choose{k_1-1}},
\end{gather*}
that is,
\[
 C_Y(2) =\frac{1}{L}
  \left( \sum_{i=3}^{P_1} k_i(i-2) +\frac{\DIS L_1 \sum_{m=1}^{k_1} (m-1){{L_1+k_1-m-2}\choose{k_1-m}}}{|{\cal B}_1|}
  +\frac{\DIS 2\hat{k}_2 {{L_1+k_1-2}\choose{k_1-1}}}{|{\cal B}_1|} \right).
\]
When we put $n=L_1+k_1-2$ and $l=L_1$, \eqref{binomialidentity} turns into
\[
{L_1+k_1-2 \choose L_1}=\sum_{m=0}^{k_1-2}(m+1){L_1+k_1-2-2-m \choose k_1-2-m}
=\sum_{m=1}^{k_1}(m-1){L_1+k_1-2-m \choose k_1-m}.
\]
Since $\DIS |{\cal B}_1|={{L_1+k_1-1}\choose{k_1}}$, we obtain \eqref{cy2}.
Equation~\eqref{cy21} follows from \eqref{cy1formula}.
\end{proof}

\subsection[Correlation function $C_Y(3)$]{Correlation function $\boldsymbol{C_Y(3)}$}
The evaluation of $C_Y(3)$ can be done in analogous way to that of $C_Y(2)$, although we have to consider various patterns so that the expression may become fairly complicated.

Using $10$-insertion and the Young diagram $Y_2$ def\/ined in the previous section, we have
\begin{gather*}
C_Y(3)  =\frac{1}{L|{\cal B}_1||{\cal B}_2||\Omega_{Y_2}|} \sum_{B_1\in {\cal B}_1} \sum_{B_2\in {\cal B}_2} \sum_{f_2\in \Omega_{Y_2}}\\
\phantom{C_Y(3)  =}{} \sum_{n=1}^L
\big(I(B_1)I(B_2)f_2\big)(n)  \big(I(B_1)I(B_2)f_2\big)(n+3),
\end{gather*}
where ${\cal B}_2 =\big\{\,B_2\in (\Z_{\ge0})^{L_2}\,\big|\,\sum\limits_{i=1}^{L_2}b_i =k_2,\,B_2=(b_1,b_2,\ldots,b_{L_2})\,\big\}$.
We call a $1**1$ pattern for an array of 4 elements of the form $1**1$ where $**$ is $00$ or $01$ or $10$ or $11$.
The following lemmas are easily understood by the def\/inition of $10$-insertion and Proposition~\ref{Prop1}.

\begin{lemma}
\label{lemma4}
In a state $f=I_1f_1$ $(f_1\in \Omega_{Y_1})$, no pattern with $1101$ and $1011$ exists.
\end{lemma}

\begin{lemma}
\label{lemma5}
For $B_2\in {\cal B}_2$ and $f_2\in \Omega_{Y_2}$, the number of the $1*1$ patterns in $f_1:=I(B_2)f_2$ is equal to that of the patterns  of the form $1001$ and $1111$ in $I_1 f_1$, and is given by
\begin{equation*}
\sum_{i=4}^{P_1} k_i(i-3) +j(B_2),
\end{equation*}
where
\[
j(B_2):=\sum_{i=1}^{L_2} \Phi_0(\tilde{b}_i;f_2(i),f_2(i+1))\qquad \big(B_2=(\tilde{b}_1,\tilde{b}_2,\ldots,\tilde{b}_{L_2})\in{\cal B}_2\big).
\]
\end{lemma}

\begin{lemma}
\label{lemma6}
Let  $\widetilde{{\cal N}}_2:=\sum\limits_{i=4}^{P_1}k_i(i-3)$\enskip
and $\widetilde{\cal R}_2 :={\cal R}_2|_{(f_1,{\cal B}_1)\to(f_2,{\cal B}_2)}
=\frac{1}{|{\cal B}_2|} \sum\limits_{B_2\in{\cal B}_2} j(B_2)$.
Then we have
\begin{equation*}
C_{Y_1}(2)=\frac{1}{L_1}\big( \widetilde{{\cal N}}_2+\widetilde{{\cal R}}_2  \big).
%\label{calN3}
\end{equation*}
\end{lemma}

Let us def\/ine
\[
{\cal N}_3:=\frac{1}{|{\cal B}_1||{\cal B}_2||\Omega_{Y_2}|}\sum_{B_1\in {\cal B}_1} \sum_{B_2\in {\cal B}_2} \sum_{f_2\in \Omega_{Y_2}}\sharp \big\{ \mbox{`1111',`1001'} \in I_1I(B_2)f_2 \big\}.
\]
From Lemmas~\ref{lemma4}--\ref{lemma6}, we f\/ind $\DIS {\cal N}_3=L_1C_{Y_1}(2)$.
If we def\/ine $\DIS {\cal R}_3$ by
\[
C_Y(3)=\frac{1}{L}\left( {\cal N}_3+{\cal R}_3\right),
\]
it denotes the term which arises from $10$-insertion $I(B_1)$, that is,
\begin{gather*}
{\cal R}_3 =\frac{1}{|{\cal B}_1||{\cal B}_2||\Omega_{Y_2}|}
  \sum_{B_1\in {\cal B}_1} \sum_{B_2\in {\cal B}_2} \sum_{f_2\in \Omega_{Y_2}}
\Big( \sharp \big\{ \mbox{`1101', `1011'} \in I(B_1)I(B_2)f_2 \big\} \\
\phantom{{\cal R}_3 =}{}
-\sharp \big\{ \mbox{`1001', `1111'} \in I_1I(B_2)f_2 \ \mbox{disappear in $I(B_1)I(B_2)f_2$} \big\} \Big).
\end{gather*}
Then we obtain the following proposition which gives a key formula to calculate $C_Y(3)$.
\begin{proposition}
\label{Prop2}
Let us define the functions;
$\Phi_1:\Z_{\ge0}\times (\F_2)^4 \to \{0,1,2\}$:
\begin{gather}
   \Phi_1(b;u_1,u_2,u_3,u_4)
 :=\left\{ \begin{array}{ll}
 2 & \big(b\ne0,\ (u_1,u_2,u_3,u_4)=(0,1,1,0),(1,0,0,1)\big), \\[1mm]
 1 & \left( b\ne0,\ \begin{array}{l} (u_1,u_2,u_3,u_4)=(1,0,0,0),(0,1,0,0),\\ (1,1,0,0),(0,0,1,0),(1,0,1,0),(1,1,1,0),\\
     (0,0,0,1),(0,1,0,1),(1,1,0,1),(0,0,1,1),\\ (1,0,1,1),(0,1,1,1)  \end{array} \right), \\[1mm]
 0 & \mbox{otherwise},
 \end{array} \right. \!\!\!\!\!
\label{capphi1}
\end{gather}
$\Phi_2:(\Z_{\ge0})^2\times (\F_2)^4 \to \{-1,0,1\}$:
\begin{gather}
   \Phi_2(b,b';u_1,u_2,u_3,u_4)
  :=\left\{\!\! \begin{array}{rl}
 1 & \big(bb'\ne0,\ (u_1,u_2,u_3,u_4)=(1,0,0,0),(0,0,0,0)\big), \\[1mm]
 -1 & \left( bb'\ne0,\; \begin{array}{l} (u_1,u_2,u_3,u_4)=(1,1,0,1),(0,1,0,1),\\ (1,1,1,1) \end{array} \!\right), \\[1mm]
 0 & \mbox{otherwise},
 \end{array} \right.\!\!\!\!\!\!
\label{capphi2}
\end{gather}
and $\Phi_3:(\Z_{\ge0})^2\times (\F_2)^4 \to \{-1,0\}$:
\begin{equation}
 \Phi_3(b,b';u_1,u_2,u_3,u_4) :=\left\{ \begin{array}{rl}
 -1 & \big(bb'\ne0,\ (u_1,u_2,u_3,u_4)=(1,1,1,1)\big), \\[1mm]
 0 & \mbox{otherwise}.
 \end{array} \right.
\label{capphi3}
\end{equation}
Then we have
\begin{gather}
C_Y(3) = \frac{1}{L}\left( {\cal N}_3+{\cal R}_3\right),
\label{Cy3eqNR} \\
{\cal N}_3  = L_1C_{Y_1}(2),
\label{calNvalue} \\
{\cal R}_3  = \frac{1}{|{\cal B}_2||{\cal B}_1||\Omega_{Y_2}|} \sum_{B_2\in {\cal B}_2}
 \sum_{B_1\in {\cal B}_1} \sum_{f_2\in \Omega_{Y_2}} \sum_{\fbm{b}\in (\F_2)^4} \notag \\
\phantom{{\cal R}_3  =}{}
 \phi_1(I(B_2)f_2;{\bm b}) \big( \Phi_1(b_1;{\bm b}) +\Phi_2(b_1,b_2;{\bm b}) +\Phi_3(b_1,b_3;{\bm b}) \big),\nonumber
%\label{CY3}
\end{gather}
where
\[
 \phi_1(f_1;{\bm b}) :=\sharp \big\{\,i\in[L_1]\,\big|\,\big(f_1(i-1),f_1(i),f_1(i+1),f_1(i+2)\big)=\bm{b}\,\big\}.
\]
\end{proposition}

\begin{proof}
Equations \eqref{Cy3eqNR} and \eqref{calNvalue} have already been shown.
From \eqref{hatr3} in Appendix~\ref{sec:app1}, we f\/ind that
\begin{gather*}
{\cal R}_3
 =\frac{1}{|{\cal B}_1||{\cal B}_2||\Omega_{Y_2}|}
 \sum_{B_2\in {\cal B}_2} \sum_{B_1\in {\cal B}_1} \sum_{f_2\in \Omega_{Y_2}} \sum_{i=1}^{L_1} \\
\phantom{{\cal R}_3=}{}
 \Big( \Phi_1(b_i;I(B_2)f_2(i-1),I(B_2)f_2(i),I(B_2)f_2(i+1),I(B_2)f_2(i+2)) \\
\phantom{{\cal R}_3=}{}
 +\Phi_2(b_{i},b_{i+1};I(B_2)f_2(i-1),I(B_2)f_2(i),I(B_2)f_2(i+1),I(B_2)f_2(i+2)) \\
\phantom{{\cal R}_3=}{}
 +\Phi_3(b_{i-1},b_{i+1};I(B_2)f_2(i-1),I(B_2)f_2(i),I(B_2)f_2(i+1),I(B_2)f_2(i+2)) \Big).
\end{gather*}
Since
\[
(b_1,b_2,\ldots,b_{L_1}) \in {\cal{B}}_1 \ \longrightarrow \ (b_i,b_i+1,\ldots,b_{L_1},b_1,\ldots,b_{i-1}) \in {\cal{B}}_1,
\]
we can rewrite
\begin{gather*}
 \sum_{B_1\in {\cal B}_1} \Phi_1(b_i;I(B_2)f_2(i-1),I(B_2)f_2(i),I(B_2)f_2(i+1),I(B_2)f_2(i+2)) \\
\qquad{} =\sum_{B_1\in {\cal B}_1} \Phi_1(b_1;I(B_2)f_2(i-1),I(B_2)f_2(i),I(B_2)f_2(i+1),I(B_2)f_2(i+2)).
\end{gather*}
Furthermore
\begin{gather*}
 \sum_{i=1}^{L_1} \Phi_1(b_1;I(B_2)f_2(i-1),I(B_2)f_2(i),I(B_2)f_2(i+1),I(B_2)f_2(i+2)) \\
\qquad{} =\sum_{\fbm{b}\in (\F_2)^4} \phi_1(I(B_2)f_2;{\bm b}) \Phi_1(b_1;{\bm b}).
\end{gather*}
Similar relations hold for $\Phi_2$ and $\Phi_3$, and we obtain
\begin{gather*}
{\cal R}_3
 =\frac{1}{|{\cal B}_1||{\cal B}_2||\Omega_{Y_2}|}
 \sum_{B_2\in {\cal B}_2} \sum_{B_1\in {\cal B}_1} \sum_{f_2\in \Omega_{Y_2}} \sum_{\fbm{b}\in (\F_2)^4} \\
\phantom{{\cal R}_3=}{}
 \phi_1(I(B_2)f_2;{\bm b}) \big( \Phi_1(b_1;{\bm b}) +\Phi_2(b_1,b_2;{\bm b}) +\Phi_3(b_1,b_3;{\bm b}) \big).\tag*{\qed}
\end{gather*}
  \renewcommand{\qed}{}
\end{proof}

\begin{theorem}
\label{theorem3}
\begin{gather}
 C_Y(3)  = \frac{1}{L}\Bigg( L_1C_{Y_1}(2) +2K_1^{(1)}\big( 3\hat{k}_3 +2k_2 -\widetilde{\cal R}_2 \big) \nonumber \\
\phantom{C_Y(3)  =}{}
  +K_1^{(1)}K_2^{(1)}\big\{ L_1 -4\hat{k}_3 -3k_2 -\widetilde{\cal R}_2 -\big( K_1^{(2)}+1 \big) L_2C_{Y_2}(1) -L_1C_{Y_1}(1,2,3) \big\} \nonumber \\
\phantom{C_Y(3)  =}{}
  -K_1^{(1)}K_2^{(1)}K_3^{(1)}\frac{(L_1-1)L_1C_{Y_1}(1,2,3)}{k_1-2} \Bigg),
\label{cy3}
\end{gather}
where
\[
K_i^{(j)} :=\frac{k_j-i+1}{L_j+k_j-i},
\]
and
\[
\widetilde{\cal R}_2 =\frac{2\hat{k}_3k_2}{L_2+k_2-1} +\frac{k_2(k_2-1)}{L_2+k_2-2}.
\]
\end{theorem}

\begin{proof}
From \eqref{RAB2} and \eqref{Lateruse} in Appendix~\ref{sec:app2}, we obtain the following equations:
\begin{gather}
 \sum_{B_1\in {\cal B}_1}  \sum_{\fbm{b}\in (\F_2)^4}
 \phi_1(I(B_2)f_2;{\bm b})  \Phi_1(b_1;{\bm b}) ={{L_1+k_1-2}\choose{k_1-1}} \left\{ 2\left( 3\hat{k}_3 +2k_2 -j(B_2) \right) \right\},
\label{phi1}
\\
  \sum_{B_1\in {\cal B}_1}
 \Big( \phi_1(I(B_2)f_2;1,0,0,0)  \Phi_2(b_1,b_2;1,0,0,0)
  +\phi_1(I(B_2)f_2;0,0,0,0)  \Phi_2(b_1,b_2;0,0,0,0) \nonumber \\
 \qquad {}
  +\phi_1(I(B_2)f_2;1,1,0,1)  \Phi_2(b_1,b_2;1,1,0,1)
  +\phi_1(I(B_2)f_2;0,1,0,1)  \Phi_2(b_1,b_2;0,1,0,1) \Big) \nonumber \\
 \qquad ={{L_1+k_1-3}\choose{k_1-2}}
 \Big( \phi_1(I(B_2)f_2;1,0,0,0) +\phi_1(I(B_2)f_2;0,0,0,0) \nonumber \\
\qquad{}
 -\phi_1(I(B_2)f_2;1,1,0,1) -\phi_1(I(B_2)f_2;0,1,0,1) \Big) \label{phi2}\\
 \quad ={{L_1+k_1-3}\choose{k_1-2}}\Bigg( L_1 -4\hat{k}_3 -3k_2 -j(B_2)
 -\sum_{i=1}^{L_2} \zeta(\tilde{b}_i;f_2(i),f_2(i+1)) -\sum_{i=4}^{P_1} k_i(i-3) \Bigg),\nonumber
\\
  \sum_{B_1\in {\cal B}_1}
 \phi_1(I(B_2)f_2;1,1,1,1)  \big( \Phi_2(b_1,b_2;1,1,1,1) +\Phi_3(b_1,b_3;1,1,1,1) \big) \nonumber \\
 \quad =-\left\{ {{L_1+k_1-3}\choose{k_1-2}} +{{L_1+k_1-4}\choose{k_1-2}} \right\} \phi_1(I(B_2)f_2;1,1,1,1),
\label{phi3}
\end{gather}
where $(\tilde{b}_1,\tilde{b}_2,\ldots,\tilde{b}_{L_2})=B_2\in{\cal B}_2$ and
\begin{gather}
 \zeta(b;u_1,u_2) :=\left\{ \begin{array}{ll}
 1 & \big(b\ne0,\ (u_1,u_2)=(1,1)\big), \\[1mm]
 0 & \mbox{otherwise}.
 \end{array} \right.
\label{zeta}
\end{gather}
Note that, from the def\/inition,
\[
\frac{1}{|{\cal B}_2||\Omega_{Y_2}|} \sum_{B_2\in{\cal B}_2} \sum_{f_2\in \Omega_{Y_2}} \phi_1(I(B_2)f_2;1,1,1,1) =L_1C_{Y_1}(1,2,3).
\]
Since
\[
(\tilde{b}_1,\tilde{b}_2,\ldots,\tilde{b}_{L_2})\in {\cal{B}}_2  \ \longrightarrow \ (\tilde{b}_i,\tilde{b}_{i+1},\ldots,\tilde{b}_{L_2},\tilde{b}_1,\ldots,\tilde{b}_{i-1})\in {\cal{B}}_2
\]
and
\begin{gather*}
\sharp \{\mbox{`11' in } f_2\in\Omega_{Y_2}\} =\sharp \{\mbox{`11' in } f_2'\in\Omega_{Y_2}\},\\
\sum_{B_2\in{\cal B}_2} \! \sum_{i=1}^{L_2} \zeta(\tilde{b}_i;f_2(i),f_2(i+1))  =\!\sum_{B_2\in{\cal B}_2}\! \zeta(\tilde{b}_1;1,1) \!\left( \sum_{i=4}^{P_1} k_i(i-3)\! \right)  ={{L_2+k_2-2}\choose{k_2-1}} L_2C_{Y_2}(1).
\end{gather*}
Hence, we have
\begin{gather*}
 {\cal R}_3  = \frac{1}{|{\cal B}_2|} \sum_{B_2\in{\cal B}_2}
  \Bigg\{ 2\left( 3\hat{k}_3 +2k_2 -j(B_2) \right) \frac{{{L_1+k_1-2}\choose{k_1-1}}}{{{L_1+k_1-1}\choose{k_1}}} \\
\phantom{{\cal R}_3  =}{}
  +\Bigg( L_1 -4\hat{k}_3 -3k_2 -j(B_2) -\Big(\zeta(\tilde{b}_1;1,1) +1\Big)\sum_{i=4}^{P_1} k_i(i-3) \Bigg) \frac{{{L_1+k_1-3}\choose{k_1-2}}}{{{L_1+k_1-1}\choose{k_1}}} \\
\phantom{{\cal R}_3  =}{}
  -\frac{{{L_1+k_1-3}\choose{k_1-2}} +{{L_1+k_1-4}\choose{k_1-2}}}{{{L_1+k_1-1}\choose{k_1}}}
   \frac{1}{|\Omega_{Y_2}|} \sum_{f_2\in \Omega_{Y_2}} \phi_1(I(B_2)f_2;1,1,1,1) \Bigg\} \\
\phantom{{\cal R}_3 }{}
=\frac{2k_1\left( 3\hat{k}_3 +2k_2 -\widetilde{\cal R}_2 \right)}{L_1+k_1-1} \\
\phantom{{\cal R}_3  =}{}
  +\frac{k_1(k_1-1)\left\{ L_1 -4\hat{k}_3 -3k_2 -\widetilde{\cal R}_2 -\left( \frac{k_2}{L_2+k_2-1}+1 \right) L_2C_{Y_2}(1) -L_1C_{Y_1}(1,2,3) \right\}}{(L_1+k_1-1)(L_1+k_1-2)} \\
\phantom{{\cal R}_3  =}{}
  -\frac{k_1(k_1-1)(L_1-1)L_1C_{Y_1}(1,2,3)}{(L_1+k_1-1)(L_1+k_1-2)(L_1+k_1-3)}.
\end{gather*}
From Proposition~\ref{Prop2}, this equation and $K_i^{(j)}$ give \eqref{cy3}.
\end{proof}

\section{Concluding remarks}
\label{sec:summ}

In this article, we showed a recursion formula for the $N$-point functions $C_Y(1,2,\ldots,N-1)$.
\mbox{Using} the formula, we explicitly calculated the second and the third nearest neighbor correlation functions.
In these estimations, we found that $C_Y(3)$ is essentially obtained from $C_Y(1)$, $C_Y(2)$ and $C_Y(1,2,3)$.
We expect that such recurrence formulae may exist for general $N$-point functions.
Obtaining the recurrence formulae and to clarify their relation to
correlation functions for quantum integrable systems are problems that will be addressed in the future.

\appendix
\section[Evaluation of ${\cal R}_3$ in  $C_Y(3)$]{Evaluation of $\boldsymbol{{\cal R}_3}$ in  $\boldsymbol{C_Y(3)}$}
\label{sec:app1}

Let $B_2 \in {\cal B}_2$, $f_2 \in \Omega_{Y_2}$ and
\begin{gather*}
\hat{R}(B_2;f_2)  :=\frac{1}{|{\cal B}_1|}\sum_{B_1 \in {\cal B}_1} \Big( \sharp \big\{ \mbox{`1101', `1011'} \in I(B_1)I(B_2)f_2\big\} \\
\phantom{\hat{R}(B_2;f_2)  :=}{}
-\sharp \big\{ \mbox{`1001', `1111'} \in I_1I(B_2)f_2\ \mbox{disappear in}\ I(B_1)I(B_2)f_2 \big\} \Big).
\end{gather*}
To evaluate $\DIS \hat{R}(B_2;f_2)$ we have to count the balance of the increment and the decrement of the $1**1$ patterns by $10$-insertion.
There are two types of such $1**1$ patterns:
\begin{description}\itemsep=0pt
\item[(A)] One of the $1$s at the ends of a $1**1$ pattern originally belong to $\DIS I_1I(B_2)f_2$;
\item[(B)] Both of the $1$s at the ends come from $10$-insertion $I_2(B_1)$.
\end{description}
In each case, the variation is listed as follows.
\begin{description}\itemsep=0pt
\item[(A)]
Let $f_1 =I(B_2)f_2$. When a block $\underbrace{\underline{10}\,\underline{10}\cdots\underline{10}}_{b_n\ne 0}$ is inserted between $f_1(n)$ and $f_1(n+1)$, the number of $1**1$ patterns of type (A) is given as:
\begin{description}\itemsep=0pt
\item[(1)]\ $f_1(n)f_1(n+1)={\bf 00}:$\\
\begin{tabular}{llll}
(a) & $0{\bf 00}0$ & $\longrightarrow\ 0{\bf 0}\underline{10}\,\underline{10}\cdots\underline{10}{\bf 0}0$ & $\Rightarrow\ \pm0\ (+0,-0),$ \\
(b) & $1\fbox{\hskip-1.2mm 10\hskip-1.1mm}{\bf 00}0$ & $\longrightarrow\ 1\fbox{\hskip-1.2mm 10\hskip-1.1mm}{\bf 0}\underline{10}\,\underline{10}\cdots\underline{10}{\bf 0}0$ & $\Rightarrow\ +1\ (+1,-0),$ \\
(c) & $0{\bf 00}1$ & $\longrightarrow\ 0{\bf 0}\underline{10}\,\underline{10}\cdots\underline{10}{\bf 0}1$ & $\Rightarrow\ +1\ (+1,-0),$ \\
(d) & $1\fbox{\hskip-1.2mm 10\hskip-1.1mm}{\bf 00}1$ & $\longrightarrow\ 1\fbox{\hskip-1.2mm 10\hskip-1.1mm}{\bf 0}\underline{10}\,\underline{10}\cdots\underline{10}{\bf 0}1$ & $\Rightarrow\ +2\ (+2,-0);$ \\
\end{tabular}
\item[(2)]\ $f_1(n)f_1(n+1)={\bf 10}:$\\
\begin{tabular}{llll}
(a) & ${\bf 1}\fbox{\hskip-1.2mm 10\hskip-1.1mm}{\bf 0}0$ & $\longrightarrow\ {\bf 1}\fbox{\hskip-1.2mm 10\hskip-1.1mm}\underline{10}\,\underline{10}\cdots\underline{10}{\bf 0}0$ & $\Rightarrow\ +1\ (+1,-0),$ \\
(b) & ${\bf 1}\fbox{\hskip-1.2mm 10\hskip-1.1mm}{\bf 0}1$ & $\longrightarrow\ {\bf 1}\fbox{\hskip-1.2mm 10\hskip-1.1mm}\underline{10}\,\underline{10}\cdots\underline{10}{\bf 0}1$ & $\Rightarrow\ +1\ (+2,-1);$ \\
\end{tabular}
\item[(3)]\ $f_1(n)f_1(n+1)={\bf 01}:$\\
\begin{tabular}{llll}
(a) & $0{\bf 01}1$ & $\longrightarrow\ 0{\bf 0}\underline{10}\,\underline{10}\cdots\underline{10}{\bf 1}1$ & $\Rightarrow\ +1\ (+1,-0),$ \\
(b) & $1\fbox{\hskip-1.2mm 10\hskip-1.1mm}{\bf 01}1$ & $\longrightarrow\ 1\fbox{\hskip-1.2mm 10\hskip-1.1mm}{\bf 0}\underline{10}\,\underline{10}\cdots\underline{10}{\bf 1}1$ & $\Rightarrow\ +1\ (+2,-1);$ \\
\end{tabular}
\item[(4)]\ $f_1(n)f_1(n+1)={\bf 11}:$\\
\begin{tabular}{llll}
(a) & $0{\bf 11}\fbox{\hskip-1.2mm 10\hskip-1.1mm}0$ & $\longrightarrow\ 0{\bf 1}\underline{10}\,\underline{10}\cdots\underline{10}{\bf 1}\fbox{\hskip-1.2mm 10\hskip-1.1mm}0$ & $\Rightarrow\ +2\ (+2,-0),$ \\
(b) & $01{\bf 11}\fbox{\hskip-1.2mm 10\hskip-1.1mm}0$ & $\longrightarrow\ 01{\bf 1}\underline{10}\,\underline{10}\cdots\underline{10}{\bf 1}\fbox{\hskip-1.2mm 10\hskip-1.1mm}0$ & $\Rightarrow\ +1\ (+2,-1),$ \\
(b$'$) & $11{\bf 11}\fbox{\hskip-1.2mm 10\hskip-1.1mm}0$ & $\longrightarrow\ 11{\bf 1}\underline{10}\,\underline{10}\cdots\underline{10}{\bf 1}\fbox{\hskip-1.2mm 10\hskip-1.1mm}0$ & $\Rightarrow\ +1\ (+3,-2),$ \\
(c) & $0{\bf 11}1$ & $\longrightarrow\ 0{\bf 1}\underline{10}\,\underline{10}\cdots\underline{10}{\bf 1}1$ & $\Rightarrow\ +1\ (+2,-1),$ \\
(d) & $01{\bf 11}11$ & $\longrightarrow\ 01{\bf 1}\underline{10}\,\underline{10}\cdots\underline{10}{\bf 1}11$ & $\Rightarrow\ \pm0\ (+2,-2),$ \\
(d$'$) & $11{\bf 11}11$ & $\longrightarrow\ 11{\bf 1}\underline{10}\,\underline{10}\cdots\underline{10}{\bf 1}11$ & $\Rightarrow\ \pm0\ (+3,-3)$,\\
\end{tabular}
\end{description}
where $+1$ $(+1,-0)$ etc. denote the numbers of the total (created, disappeared) $1**1$ patterns.
\end{description}
Note that total number does not depend on the length of the blocks and that the variation of $1**1$ patterns is determined for the four elements $f_1(n-2)f_1(n-1)f_1(n)f_1(n+1)$.
Hence, for $b_n\ne0$, the contribution from a boundary and a block is summarized as
\[
\left\{ \begin{array}{ll}
 0 & \big((0,0,0,0),(1,1,1,1)\big), \\[1mm]
 1 & \left( \begin{array}{l} (1,0,0,0),(0,1,0,0),(1,1,0,0),(0,0,1,0),\\ (1,0,1,0),(1,1,1,0),
     (0,0,0,1),(0,1,0,1),\\ (1,1,0,1),(0,0,1,1),(1,0,1,1),(0,1,1,1)\\ \end{array} \right), \\[1mm]
 2 & \big((0,1,1,0),(1,0,0,1)\big). \\
 \end{array} \right.
\]
For example, there is no contribution for $\big(f_1(n-1),f_1(n),f_1(n+1),f_1(n+2)\big)=(0,0,0,0)$, $(1,1,1,1)$,
and $+2$ for  $\big(f_1(n-1),f_1(n),f_1(n+1),f_1(n+2)\big)=(0,1,1,0)$,  $(1,0,0,1)$.

\begin{description}\itemsep=0pt
\item[(B)] When blocks are inserted between $f_1(n-1)$ and $f_1(n)$ and between $f_1(n)$ and $f_1(n+1)$
($b_{n-1}\ne0,\ b_n\ne0$), the total number of $1**1$ patterns generated between two blocks is given as:
\begin{description}\itemsep=0pt
\item[(1)]\ $f_1(n-1)f_1(n)f_1(n+1)={\bf 000}:$\\[1mm]
$\longrightarrow\ {\bf 0}\underline{10}\cdots\underline{10}{\bf 0}\underline{10}\,\underline{10}\cdots\underline{10}{\bf 0}\ \Rightarrow\ (+1,\pm0),$
\item[(2)]\ $f_1(n-1)f_1(n)f_1(n+1)={\bf 001}:$\\[1mm]
$\longrightarrow\ {\bf 0}\underline{10}\cdots\underline{10}{\bf 0}\underline{10}\,\underline{10}\cdots\underline{10}{\bf 1}\ \Rightarrow\ (+1,-1),$
\item[(3)]\ $f_1(n-1)f_1(n)f_1(n+1)={\bf 010}:$\\[1mm]
$\longrightarrow\ {\bf 0}\underline{10}\cdots\underline{10}{\bf 1}\fbox{\hskip-1.2mm 10\hskip-1.1mm}\underline{10}\,\underline{10}\cdots\underline{10}{\bf 0}\ \Rightarrow\ (\pm0,\pm0),$
\item[(4)]\ $f_1(n-1)f_1(n)f_1(n+1)={\bf 011}:$\\[1mm]
$\longrightarrow\ {\bf 0}\underline{10}\cdots\underline{10}{\bf 1}\underline{10}\,\underline{10}\cdots\underline{10}{\bf 1}\ \Rightarrow\ (+1,-1),$
\item[(5)]\ $f_1(n-1)f_1(n)f_1(n+1)={\bf 100}:$\\[1mm]
$\longrightarrow\ {\bf 1}\fbox{\hskip-1.2mm 10\hskip-1.1mm}\underline{10}\cdots\underline{10}{\bf 0}\underline{10}\,\underline{10}\cdots\underline{10}{\bf 0}\ \Rightarrow\ (+1,-1),$
\item[(6)]\ $f_1(n-1)f_1(n)f_1(n+1)={\bf 101}:$\\[1mm]
$\longrightarrow\ {\bf 1}\fbox{\hskip-1.2mm 10\hskip-1.1mm}\underline{10}\cdots\underline{10}{\bf 0}\underline{10}\,\underline{10}\cdots\underline{10}{\bf 1}\ \Rightarrow\ (+1,-2),$
\item[(7)]\ $f_1(n-1)f_1(n)f_1(n+1)={\bf 110}:$\\[1mm]
$\longrightarrow\ {\bf 1}\underline{10}\cdots\underline{10}{\bf 1}\fbox{\hskip-1.2mm 10\hskip-1.1mm}\underline{10}\,\underline{10}\cdots\underline{10}{\bf 0}\ \Rightarrow\ (\pm0,\pm0),$
\item[(8)]\ $f_1(n-1)f_1(n)f_1(n+1)={\bf 111}:$\\[1mm]
\begin{tabular}{ll}
(a) & $\longrightarrow\ 0{\bf 1}\underline{10}\cdots\underline{10}{\bf 1}\underline{10}\,\underline{10}\cdots\underline{10}{\bf 1}\ \Rightarrow\ (+1,-1),$ \\
(b) & $\longrightarrow\ 1{\bf 1}\underline{10}\cdots\underline{10}{\bf 1}\underline{10}\,\underline{10}\cdots\underline{10}{\bf 1}\ \Rightarrow\ (+1,-2);$ \\
\end{tabular}
\end{description}
Here $(+1,\pm0)$ etc. denote the number of $1**1$ patterns which (arise between blocks, are eliminated by the $10$-insertion of blocks).

When a block is inserted between $f_1(n-2)$ and $f_1(n-1)$ and another block is inserted between $f_1(n)$ and $f_1(n+1)$;
\begin{description}\itemsep=0pt
\item[(9)]\ $f_1(n-1)f_1(n)f_1(n+1)={\bf 111}:$\\[1mm]
\begin{tabular}{ll}
(a) & $\longrightarrow\ 0\underline{10}\cdots\underline{10}{\bf 11}\underline{10}\,\underline{10}\cdots\underline{10}{\bf 1}\ \Rightarrow\ (\pm0,\pm0),$ \\
(b) & $\longrightarrow\ 1\underline{10}\cdots\underline{10}{\bf 11}\underline{10}\,\underline{10}\cdots\underline{10}{\bf 1}\ \Rightarrow\ (\pm0,-1)$. \\
\end{tabular}
\end{description}
\end{description}
From these list, we f\/ind that the pattern $\DIS f_1(n-1)f_1(n)f_1(n+1)f_1(n+2)$ determines the variation locally.
For $b_n b_{n+1} \ne 0$,
\[
\left\{ \begin{array}{rl}
 1 & \big((1,0,0,0),(0,0,0,0)\big), \\[1mm]
 -1 & \left(  \begin{array}{l} (1,1,0,1),(0,1,0,1),\\ (1,1,1,1)\\ \end{array} \right), \\[1mm]
 0 & \mbox{otherwise},
 \end{array} \right.
\]
and  for $b_{n-1}\ne 0,\ b_n=0,\ b_{n+1}\ne 0$,
\[
\left\{ \begin{array}{rl}
 -1 & \ (1,1,1,1), \\[1mm]
 0 & \ \mbox{otherwise}.
 \end{array} \right.
\]

Therefore, using the functions $\Phi_1$, $\Phi_2$ and $\Phi_3$ \eqref{capphi1}--\eqref{capphi3} in Proposition~\ref{Prop2},
we obtain
\begin{gather}
\hat{R}(B_2;f_2) =\frac{1}{|{\cal B}_1|}\sum_{B_1\in {\cal B}_1}\sum_{i=1}^{L_1}  \notag\\
\phantom{\hat{R}(B_2;f_2) =}{}
  \Big( \Phi_1(b_i;I(B_2)f_2(i-1),I(B_2)f_2(i),I(B_2)f_2(i+1),I(B_2)f_2(i+2)) \notag\\
\phantom{\hat{R}(B_2;f_2) =}{}
 +\Phi_2(b_{i},b_{i+1};I(B_2)f_2(i-1),I(B_2)f_2(i),I(B_2)f_2(i+1),I(B_2)f_2(i+2)) \label{hatr3}\\
\phantom{\hat{R}(B_2;f_2) =}{}
 +\Phi_3(b_{i-1},b_{i+1};I(B_2)f_2(i-1),I(B_2)f_2(i),I(B_2)f_2(i+1),I(B_2)f_2(i+2)) \Big).\notag
\end{gather}

\section{Derivation of equations~(\ref{phi1}), (\ref{phi2}) and (\ref{phi3})}
\label{sec:app2}

From \eqref{hatr3} and Proposition~\ref{Prop2}, we f\/ind
\begin{gather*}
\hat{R}(B_2;f_2) =\frac{1}{|{\cal B}_1|}\sum_{B_1\in {\cal B}_1}\sum_{\fbm{b}\in (\F_2)^4}
  \phi_1(I(B_2)f_2;{\bm b}) \big( \Phi_1(b_1;{\bm b}) +\Phi_2(b_1,b_2;{\bm b}) +\Phi_3(b_1,b_3;{\bm b}) \big).
\end{gather*}
Let $f_1:=I(B_2)f_2$. First we consider the term:
\begin{gather*}
\hat{R}_A(B_2;f_2) :=\frac{1}{|{\cal B}_1|}\sum_{B_1\in {\cal B}_1}\sum_{\fbm{b}\in (\F_2)^4}
\phi_1(I(B_2)f_2;{\bm b}) \Phi_1(b_1;{\bm b})\\
 \phantom{\hat{R}_A(B_2;f_2) \;}{} =\frac{1}{|{\cal B}_1|}\sum_{B_1\in {\cal B}_1}\sum_{n=1}^{L_1}\Phi_1(b_1;f_1(n-1),f_1(n),f_1(n+1),f_1(n+2)).
\end{gather*}
For $b \ne 0$, we observe $\Phi_1(b;{\bf b}) =\sharp \big\{ \mbox{`10',`01'}\in{\bf b} \big\} -\sharp \big\{ \mbox{`101',`010'}\in{\bf b} \big\}$.
There are $\hat{k}_2$ `10's, the same number of `01's and $j(B_2) +\sharp \big\{ \mbox{`\underline{10}' inserted between consecutive 1's} \big\}$ `101's  in $f_1$.
Since the number of `010's is equal to $\sharp \big\{ \mbox{`\underline{10}'} \big\} -\sharp \big\{ \mbox{`\underline{10}' inserted between consecutive 1's} \big\}$, we f\/ind
\begin{equation}
\hat{R}_A(B_2;f_2)=6\hat{k}_2-2j(B_2)-2k_2.
\label{RAB2}
\end{equation}

The rest terms can be evaluated by counting the numbers of 3-tuples in $f_1$.
\begin{itemize}\itemsep=0pt
\item the number of $\big(f_1(n-1),f_1(n),f_1(n+1)\big) =(0,0,1),\,(1,0,0)$ is
\[
2\hat{k}_2 -\sharp \big\{ n\in[L_2] \big| (f_1(n-1),f_1(n),f_1(n+1))=(1,0,1) \big\};
\]
\item the number of $\big(f_1(n-1),f_1(n),f_1(n+1)\big) =(0,1,0)$ is
\[
k_2 -\sum_{i=1}^{L_2} \zeta(\tilde{b}_i;f_2(i),f_2(i+1))
\]
where $\zeta(b;u_1,u_2)$ is def\/ined by \eqref{zeta};
\item the number of $\big(f_1(n-1),f_1(n),f_1(n+1)\big) =(0,1,1),\,(1,1,0)$ is
\[
2\hat{k}_3 +2\sum_{i=1}^{L_2} \zeta(\tilde{b}_i;f_2(i),f_2(i+1));
\]
\item the number of $\big(f_1(n-1),f_1(n),f_1(n+1)\big) =(1,0,1),\,(1,1,1)$ is
\[
\sum_{i=4}^{P_1} k_i(i-3) +j(B_2).
\]
\end{itemize}
Hence we have
\begin{gather*}
 \sharp\left\{ \mbox{`000'}  \in I(B_2)f_2  \right\}- \sharp\left\{ \mbox{`101'}  \in I(B_2)f_2  \right\}\\
\qquad {} =L_1 -4\hat{k}_3 -3k_2 -j(B_2) -\sum_{i=1}^{L_2} \zeta(\tilde{b}_i;f_2(i),f_2(i+1))-\sum_{i=4}^{P_1} k_i(i-3).
\end{gather*}
Therefore
\begin{gather}
 \phi_1(I(B_2)f_2;1,0,0,0) +\phi_1(I(B_2)f_2;0,0,0,0)
 -\phi_1(I(B_2)f_2;1,1,0,1)  \notag\\
\qquad{} -\phi_1(I(B_2)f_2;0,1,0,1) =\sharp\left\{ \mbox{`000'}  \in I(B_2)f_2  \right\}- \sharp\left\{ \mbox{`101'}  \in I(B_2)f_2  \right\}\notag\\
 \qquad{} =L_1 -4\hat{k}_3 -3k_2 -j(B_2) -\sum_{i=1}^{L_2} \zeta(\tilde{b}_i;f_2(i),f_2(i+1))-\sum_{i=4}^{P_1} k_i(i-3).
\label{Lateruse}
\end{gather}
From \eqref{RAB2} and \eqref{Lateruse}, $\hat{R}(B_2;f_2)$ can be obtained as shown in the proof of Theorem~\ref{theorem3}.

\subsection*{Acknowledgements}

The authors wish to thank Professor Atsushi Nagai for useful comments.

\pdfbookmark[1]{References}{ref}
\LastPageEnding

\end{document}